\documentclass[pra,aps]{revtex4-2}

\usepackage{graphicx}
\usepackage{dcolumn}
\usepackage{bm}

\usepackage{xcolor}
\usepackage{CJKutf8}
\begin{document}
\begin{CJK*}{UTF8}{gbsn}
\title{Quantum metrology timing limits of biphoton frequency comb}
\author{Baihong Li$^{1}$}
\email{li-baihong@163.com}

\author{Qi-qi Li}

\author{Boxin Yuan$^{1}$}






\author{\\ Ruifang Dong$^{2,3}$}%
 \email{dongruifang@ntsc.ac.cn}

\author{Shougang Zhang$^{2,3}$}%

\author{Rui-Bo Jin$^{4}$}%
 \email{jin@wit.edu.cn}

\affiliation{$^{1}$
School of Physics and Information Science, Shaanxi University of Science and Technology, Xi’an 710021, China
}%
\affiliation{$^{2}$
Key Laboratory of Time and Frequency Primary Standards, National Time Service Center, Chinese
Academy of Sciences, Xi’an 710600, China
}%

\affiliation{$^{3}$
School of Astronomy and Space Science, University of Chinese Academy of Sciences, Beijing 100049, China
}%

\affiliation{$^{4}$
 Hubei Key Laboratory of Optical Information and Pattern Recognition, Wuhan Institute of Technology,
Wuhan 430205, China
}%

\begin{abstract}
Biphoton frequency comb (BFC), which encompasses multiple discrete frequency modes and represents high-dimensional frequency entanglement, is crucial in quantum information processing due to its high information capacity and error resilience. It also holds significant potential for enhancing timing precision in quantum metrology. Here, we examine quantum metrology timing limits using the BFC as a probe state and derive a quantum Cramér-Rao bound that scales quadratically with the number of frequency modes. Under ideal conditions (zero loss and perfect visibility), this bound can be saturated by both spectrally non-resolved Hong-Ou-Mandel (HOM) interferometry at zero delay and spectrally resolved HOM interferometry at arbitrary delays. In particular, under imperfect experimental conditions, Fisher information rapidly increases up to its maximum as the mode number increases for a fixed time delay close to zero, indicating that increasing the mode number is an optimal strategy for improving the timing precision in practice. Furthermore, compared with spectrally non-resolved measurement, spectrally resolved measurement is a better strategy due to its higher Fisher information, shorter measurement times, and ambiguity-free dynamic range.

\end{abstract}

\maketitle


\section{\label{sec:1}INTRODUCTION}
Hong-Ou-Mandel (HOM) interference \cite{PRL1987,RPP2021,Jin2024review} is a well-known two-photon quantum interference effect that cannot be achieved in classical interference. The HOM effect has many important applications in advanced quantum technologies and, in particular, it has shown the powerful metrological potential for high precision imaging and sensing applications \cite{NATURE2001,PRL2006,NP2011}. As a result, the HOM interferometer has become a valuable apparatus for quantum parameter estimation \cite{PRL2007,SA2018,npj2019,Cyril2023,PRA2020,Fabre2020,PRA2021,Fabre2021PRA,PRA2022,CHENPRAPP2023,PRA2023,PRL2023,PRL2024,CHENPRAPP2023,PRAPP2023}.
Furthermore, analyzing the bounds on precision achievable in an estimation protocol using the Cramér-Rao bound (CRB) \cite{Cramer} and the ultimate precision achievable through the quantum Cramér-Rao bound (QCRB) \cite{Helstrom} has become a useful tool to determine the sensitivity of two-photon interferometry techniques for metrological applications. 




Biphoton frequency comb \cite{NP2019,npjQ2020}(BFC), also referred to as a “mode-locked two-photon state" \cite{PRL2003,PRA2012} or frequency-bin entanglement \cite{PRA2010,OE2019-2,PRL2020,APLP2022}, which includes multiple discrete frequency modes and represents high-dimensional frequency entanglement \cite{NRP2020}, is crucial in quantum information processing \cite{Optica2017,NP2019-} due to its high information capacity and error resilience. Moreover, the BFC has many other important applications, such as generation of high-dimensional hyperentanglement\cite{NP2015} and complex quantum states \cite{Science2016}, quantum walks \cite{SA2020}, and quantum logic \cite{NP2009}. Recently, there has been increasing attention on the applications of BFC in improving the timing precision in quantum metrology \cite{Fabre2020,Fabre2021PRA,JINOE2023}. However, to the best of our knowledge, the ultimate quantum metrology timing limits using BFC as a probe state have not yet been investigated.


In this paper, we examine the ultimate quantum metrology timing limits using BFC as a probe state and derive the QCRB that can be achieved with such a probe state. By calculating the Fisher information (FI), we determine the optimal measurement strategy and the ideal conditions necessary to saturate the QCRB. Furthermore, we analyze the ultimate quantum metrology timing limits that can be obtained under practical conditions and determine the optimal measurement strategy for rapidly approaching these limits. We also provide a comparative analysis of two measurement strategies for quantum metrology timing: spectrally non-resolved and resolved HOM interferometry. We find that the QCRB scales quadratically with the number of frequency modes. Under ideal conditions, i.e., zero loss and perfect visibility, the QCRB can be saturated by spectrally non-resolved HOM interferometry at zero delay and by spectrally resolved HOM interferometry at arbitrary delays. In practical scenarios, FI significantly increases with the increase of mode number for a fixed time delay close to zero, indicating that the use of such a state as a probe is an optimal strategy for improving the timing precision in practice. Furthermore, we show that compared to spectrally non-resolved measurement, spectrally resolved measurement is a better strategy due to its higher FI, shorter measurement time, and ambiguity-free dynamic range.

The remainder of the paper is organized as follows. In Sec. \ref{sec:2}, we derive the QCRB using BFC as a probe state.  We analyze FI of spectrally non-resolved HOM interferometry with BFC in Sec. \ref{sec:3} and spectrally resolved one in Sec. \ref{sec:4}, respectively,  and present the optimal measurement strategy to saturate the QCRB. Section \ref{sec:5} compares the results obtained from spectrally non-resolved and spectrally resolved HOM interferometry. Finally, Section \ref{sec:conclude} summarizes the results and concludes the paper.

\section{\label{sec:2} Quantum Cramér-Rao bound and sensitivity using biphoton frequency comb as a probe state}

In a typical metrological protocol, a probe is prepared in an initial state, and then undergoes a dynamical evolution depending on a parameter to be estimated. We consider the generic task of estimating an unknown time parameter $\tau$, for example, the time delay in a physical system. We prepare a probe state $|\Psi_0\rangle$, which evolves to $|\Psi(\tau)\rangle$ after interacting with the physical system. The transformed state is then subjected to a specific measurement strategy to obtain an estimator of $\tau$. There exists a fundamental limit to the precision of estimation for a pure state \cite{PRL2007, npj2019}, regardless of the details of the final measurement step
\begin{equation}
\label{delta}
\delta \tau \geq \frac{1}{\sqrt{NQ}}=\delta \tau_{QCR},
\end{equation}
where
\begin{equation}
\label{Q}
Q=4\Big(\Big\langle\frac{\partial\Psi(\tau)}{\partial\tau}\Big|\frac{\partial\Psi(\tau)}{\partial\tau}\Big\rangle-\Big|\Big\langle \Psi(\tau) \Big|\frac{\partial\Psi(\tau)}{\partial\tau}\Big\rangle\Big|^2\Big).
\end{equation}
$N$ is the number of repetitions of the experiment. This statement provides the QCRB and indicates that one will never achieve a precision better than this bound, regardless of the ingenious measurement procedure the experimenter may contrive. Since the QCRB is associated with a particular quantum state, it is clear that the most important thing is to choose an appropriate probe state.

Let us now consider the BFC as a probe state. Such a state represents a discrete multi-mode frequency-entangled state, which can be expressed as follows: \cite{PRL2003,PRA2012}
\begin{eqnarray}
\label{psi}
|\Psi_0\rangle=\sum_m\int d\Omega f(\Omega+m\mu)\hat{a}_{1}^\dag(\omega_{s0}+\Omega)\hat{a}_{2}^\dag(\omega_{i0}-\Omega)|0\rangle,
\end{eqnarray}
where $m$ is mode number representing the total number of frequency pairs. $f(\Omega)$, the joint spectral amplitude (JSA) of biphotons, gives the spectral distribution of a single mode and fulfills the normalized condition $\int d\Omega |f(\Omega)|^2$=1. $\Omega$ is the frequency detuning around the center frequency, and $\hat{a}^\dag(\omega_{s,i})$ is the creation operator, with subscripts $s$ and $i$ denoting the signal and idler photons, respectively. $|0\rangle$ stands for a vacuum state. For each frequency pair of the two-photon state, we have $\omega_p=\omega_{s}+\omega_{i}$.  If we define the mode spacing as $\mu$, then $\omega_{s}=\omega_{s0}+\Omega+m\mu/2, \omega_{i}=\omega_{i0}-\Omega-m\mu/2$, where $\omega_{s0},\omega_{i0}$ denote the central frequency of the frequency pair that is most close to the degenerate frequency. Then, we define the difference between the central frequency of the frequency pair as $\Delta=\omega_{s0}-\omega_{i0}$. If $\Delta=0$, that is two photons have the same center frequency and they are frequency-degenerate; otherwise, they are frequency-nondegenerate. In the following discussions, we take the JSA as a Gaussian function, i.e., 
$f(\Omega+m\mu)= \exp(-((2k-m-1)\mu/2+\Omega)^2/4\sigma^2)$. 

Now, we use the BFC probe state to interact with a dynamic system, e.g, a HOM interferometer. Each photon of the pair is injected into one of the two arms of a HOM interferometer, and a relative time delay $\tau$ is introduced in the path of the idler photon. If we only consider the part that contributs to coincidence counts, the quantum state after the beamspliter in a HOM scheme would be
\begin{eqnarray}
\label{psi-tau}
|\Psi(\tau)\rangle=\frac{1}{\sqrt2}\sum_m\int d\Omega f(\Omega+m\mu)\Big(e^{i(\omega_{s0}+\Omega)\tau}\hat{a}_{1}^\dag(\omega_{s0}+\Omega)\hat{a}_{2}^\dag(\omega_{i0}-\Omega)|0\rangle
+e^{i(\omega_{i0}-\Omega)\tau}\hat{a}_{1}^\dag(\omega_{i0}-\Omega)\hat{a}_{2}^\dag(\omega_{s0}+\Omega)|0\rangle\Big),
\end{eqnarray}
For this state, the QCRB on the estimation of time delays writes (see details in Appendix A)
\begin{equation}
\label{QCR}
\delta \tau_{QCR}=\frac{1}{N^{1/2}}\frac{1}{[(m^2-1)\mu^2/3+\Delta^2+4\sigma^2]^{1/2}},
\end{equation}
where $\sigma=\sqrt{\langle\Omega^2\rangle-\langle\Omega\rangle^2}$ is the RMS (root mean square) bandwidth of a single frequency mode. Eq. (\ref{QCR}) provides the ultimate limit of sensitivity that can be theoretically achieved for the state in Eq. (\ref{psi-tau}) when it is used as a probe. In other words, no experiment utilizing this state as a probe can yield better results than this theoretical limit. For $m=1$, namely a single frequency mode, the result is deduced to $1/\sqrt{N(\Delta^2+4\sigma^2)}$, and when two photons are frequency-degenerate, i.e., $\Delta=0$, it becomes $1/\sqrt{N}2\sigma$. These results are consistent with those obtained in \cite{SA2018,npj2019,CHENPRAPP2023}, which suggest that the potential use of non-degenerate and large-bandwidth frequency entanglement could serve as an alternative for enhanced resolution in HOM interferometry. On the other hand, the dependence of the QCRB on mode number $m$ and mode spacing $\mu$ in Eq. (\ref{QCR}) suggests that multi-mode frequency entanglement with large spacing could  further enhance the resolution of HOM interferometry. Since the former has been investigated, this paper mainly focuses on the effects of mode number $m$ and mode spacing $\mu$, which have not yet been explored.

In a particular measurement, the sensitivity can be obtained by measuring an observable $\hat{S}$ and calculating its mean value $\langle\hat{S}\rangle$, that depends on $\tau$, to estimate the value of the variable. The precision of this measurement can be calculated as \cite{npj2019} 
\begin{equation}
\label{sensitivity}
\delta \tau=\sqrt{\langle(\Delta\hat{S})^2\rangle}\Big/\Big|\frac{\partial\langle\hat{S}\rangle}{\partial\tau}\Big|,
\end{equation}
where  $\langle(\Delta\hat{S})^2\rangle=\langle(\hat{S})^2\rangle-\langle\hat{S}\rangle^2$. The numerator represents the variance of the operator and the denominator is the sensitivity of the mean value of the operator to the variable $\tau$. The mean value of the operator, i.e., the probability to obtain a coincidence count, considering photon loss $\gamma$ and imperfect experimental visibility $V$ \cite{SA2018,npj2019}, is (see more details in Appendix B)
\begin{eqnarray}
\label{R2}
\langle\hat{S}\rangle=\frac{1}{2}(1-\gamma)^2\Big(1-\frac{V}{m}e^{-2\sigma^2\tau^2}\cos(\Delta\tau+\phi)\frac{\sin(m\theta)}{\sin\theta}\Big).
\end{eqnarray}
where $\theta=\mu\tau$, and $\phi$ represents the relative phase between two discrete multi-mode frequency entangled states. $\phi$  is a controllable constant phase that determines the phase of the oscillating signal, and the coincidence count exhibits a dip or a peak around zero delay when the phase is zero or $\pi$, respectively \cite{PRL2009}. We set $\phi=\pi$ in the following discussions.

The derivative of the mean value $\langle\hat{S}\rangle$ is 
\begin{equation}
\frac{\partial\langle\hat{S}\rangle}{\partial\tau}=e^{-2\sigma^2\tau^2}V(1-\gamma)^2\csc(\theta)\Big(m\mu\cos(m\theta)\cos(\Delta\tau)-\{\cos(\Delta\tau)[4\sigma^2\tau+\mu\cot(\theta)]+\Delta\sin(\Delta\tau)\}\sin(m\theta)\Big)/2m,
\end{equation}

If we make use of  $\langle\hat{S}^2\rangle=\langle\hat{S}\rangle$, then the variance of the operator becomes $\langle(\Delta\hat{S})^2\rangle=\langle\hat{S}\rangle(1-\langle\hat{S}\rangle)$. Therefore, the precision of estimating $\tau$ is written as
\begin{eqnarray}
\label{deltatau}
\delta\tau=\frac{\Big(\sin^2(\theta)(1-\gamma)^2AB\Big)^{\frac{1}{2}}}{V(1-\gamma)^2\Big(m\mu\cos(m\theta)\cos(\Delta\tau)-\{\cos(\Delta\tau)[4\sigma^2\tau+\mu\cot(\theta)]+\Delta\sin(\Delta\tau)\}\sin(m\theta)\Big)},
\end{eqnarray}
with
\begin{eqnarray}
A=&&e^{2\sigma^2\tau^2}m +V\cos(\Delta\tau)\csc(\theta)\sin(m\theta),\nonumber\\
B=&&e^{2\sigma^2\tau^2}(1+2\gamma-\gamma^2) m -V(1-\gamma)^2\cos(\Delta\tau)\csc(\theta)\sin(m\theta).
\end{eqnarray}
In the ideal case, i.e., $\gamma=0,V=1$, Eq.(\ref{deltatau}) can be simplified to
\begin{eqnarray}
\label{deltatau-1}
\delta\tau_{ideal}=\frac{\{e^{4\sigma^2\tau^2} m^2 \sin(\theta)^2-\cos(\Delta\tau)^2\sin(m\theta)^2\}^{\frac{1}{2}}}{m\mu\cos(m\theta)\cos(\Delta\tau)-\{\cos(\Delta\tau)[4\sigma^2\tau+\mu\cot(\theta)]+\Delta\sin(\Delta\tau)\}\sin(m\theta)},
\end{eqnarray}
In the limit of $\tau \to 0$, we have
\begin{equation}
\label{tau-ideal}
\lim_{\tau \to 0}\delta\tau_{ideal}=\frac{1}{[(m^2-1)\mu^2/3+\Delta^2+4\sigma^2]^{\frac{1}{2}}}.
\end{equation}
It can be seen that the QCRB in Eq.(\ref{QCR}) can be recovered through HOM interferometry at zero delay, under conditions of zero loss and perfect visibility. This indicates that the measurement strategy of HOM interferometry is optimal.

Now, we can provide an intuitive understanding of why the timing precision can be improved using BFC as a probe state through HOM interferometry. The sensitivity, the smallest size that can be resolved, is determined in principle by the interferogram's minimum period. This means that the narrower the interferogram, the higher the sensitivity. From the coincidence probability in Eq.(\ref{R2}), we see that it represents a temporal HOM interference with a multi-mode frequency-entangled state. The Gaussian envelope arises from the Fourier transform of the spectral distribution of a single mode within the frequency comb, and its width is determined solely by the single-photon bandwidth $\sigma$. The cosine oscillation stems from the detuning of the center frequency between the two photons, namely $\Delta$, resulting in a spatial quantum beating \cite{PRL1988,PRL2009,npj2019,IEEE Photonics2019} with a period of $2\pi/\Delta$ within the temporal HOM interferogram. This oscillation gets faster as frequency detuning $\Delta$ increases, indicating that the temporal resolution can be significantly enhanced by increasing $\Delta$. The term $\sin(m\theta)/\sin\theta$, referred to as the "details factor" in \cite{JINOE2023}, represents a finer modulation on the temporal HOM interferogram caused by mode number $m$ and mode spacing $\mu$. This leads to narrower dips (peaks) in HOM interferogram \cite{JINOE2023},  further improving the temporal resolution as $m$ and $\mu$ increase. Consequently, the overall interferogram contains multiple replicated narrow dips or peaks \cite{PRL2003, NP2015,npj2021,OE2019,APLP2022, JINOE2023} caused by mode number $m$ and mode spacing $\mu$, as well as the beatings caused by frequency detuning $\Delta$, as shown in \cite{PRA2012}. The width of these dips or peaks  is inversely proportional to the product of the mode number $m$ and the mode spacing $\mu$. Therefore, the final resolution can be improved by increasing both mode number $m$, mode spacing $\mu$, as well as the frequency detuning $\Delta$, as indicated in Eqs. (\ref{QCR}) and (\ref{tau-ideal}). 

\section{\label{sec:3} Fisher information of Spectrally non-resolved HOM interferometry with biphoton frequency comb}
Next we must confirm that we can experimentally realize this potential benefit using an appropriate measurement strategy, i.e., one that allows us to saturate Eq. (\ref{QCR}). As we will see in the following, this can be accomplished through coincidence detection in a standard HOM interferometer. In the case of a real HOM interferometer, considering photon loss $\gamma$ and imperfect experimental visibility $V$ \cite{SA2018,npj2019}, there are three possible measurement outcomes: either both photons are detected, one photon is detected, or no photon detected, which correspond to three probability distributions
\begin{eqnarray}
\label{R123}
&&R_2(\tau)=\frac{1}{2}(1-\gamma)^2\Big(1+\frac{V}{m}e^{-2\sigma^2\tau^2}\cos(\Delta\tau)\frac{\sin(m\theta)}{\sin\theta}\Big),
\nonumber\\
&&R_1(\tau)=\frac{1}{2}(1-\gamma)^2\Big(\frac{1+3\gamma}{1-\gamma}-\frac{V}{m}e^{-2\sigma^2\tau^2}\cos(\Delta\tau)\frac{\sin(m\theta)}{\sin\theta}\Big),\nonumber\\
&&R_0(\tau)=\gamma^2.
\end{eqnarray}
where subscripts 0, 1, and 2 denote the number of detectors that register a click, corresponding to total loss, bunching and coincidence, respectively. The outcome probabilities from this measurement can now be used to construct an estimator for the value of $\tau$. An estimator $\tilde\tau$ is a function of the experimental data that allows us to infer the unknown time delay using a specific statistical model for the probability distribution of measurement outcomes. As such, it is itself a random variable that can be constructed from the probability distributions $R_i(\tau)$  as a function of the time delay.  The average of an unbiased estimator corresponds to the real-time delay. According to classical estimation theory, the standard deviation of any such estimator is lower-bounded by
\begin{equation}
\label{CR}
\delta \tau_{CR}=\frac{1}{\sqrt{NF_{\tau}}} \geq \delta \tau_{QCR},
\end{equation}
where the FI reads
\begin{equation}
\label{FI}
F_{\tau}=\sum_{i}\frac{(\partial_{\tau} R_i(\tau))^2}{R_i(\tau)}=\frac{(\partial_{\tau}R_2(\tau))^2}{R_2(\tau)}+\frac{(\partial_{\tau}R_1(\tau))^2}{R_1(\tau)}+\frac{(\partial_{\tau}R_0(\tau))^2}{R_0(\tau)}.
\end{equation}
 Substituting Eq.(\ref{R123}) into Eq.(\ref{FI}), we finally arrive at
\begin{eqnarray}
\label{FI-1}
F_{\tau}=\frac{\{V^2(1-\gamma)^2(1+\gamma) \csc^2(\theta) \} \{\Delta\sin(\Delta\tau)\sin(m\theta)+\cos(\Delta\tau)[(4\sigma^2\tau+\mu\cot(\theta))\sin(m\theta)-m\mu\cos(m\theta)]\}^2}{\{e^{2\sigma^2\tau^2}m+V\cos(\Delta\tau)\csc(\theta)\sin(m\theta)\}\{e^{2\sigma^2\tau^2}m(1+3\gamma)-V(1-\gamma)\cos(\Delta\tau)\csc(\theta)\sin(m\theta)\}},
\end{eqnarray}
In the ideal case, i.e., $\gamma=0,V=1$, Eq.(\ref{FI-1}) can be simplified to
\begin{eqnarray}
\label{FI-ideal}
F_{\tau,ideal}=\frac{\csc^2(\theta) \{m\mu\cos(\Delta\tau)\cos(m\theta)-[\cos(\Delta\tau)(4\sigma^2\tau+\mu\cot(\theta))+\Delta\sin(\Delta\tau)]\sin(m\theta)\}^2}{e^{4\sigma^2\tau^2}m-\{\cos(\Delta\tau)\csc(\theta)\sin(m\theta)\}^2},
\end{eqnarray}
This limit is known as the CRB and is associated with a particular quantum state and a specific measurement strategy. The QCRB can be obtained by maximizing over all possible measurements on the probe state. By evaluating the FI for this set of probabilities, we find that its upper bound is achieved in the ideal case, i.e., $\gamma=0,V=1$  at zero delay as, given by
\begin{equation}
\label{FI0}
\lim_{\tau \to 0}F_{\tau,ideal}=(m^2-1)\mu^2/3+\Delta^2+4\sigma^2.
\end{equation}
\begin{figure}[th]
\begin{picture}(380,200)
\put(0,0){\makebox(375,200){
\scalebox{0.58}[0.58]{
\includegraphics{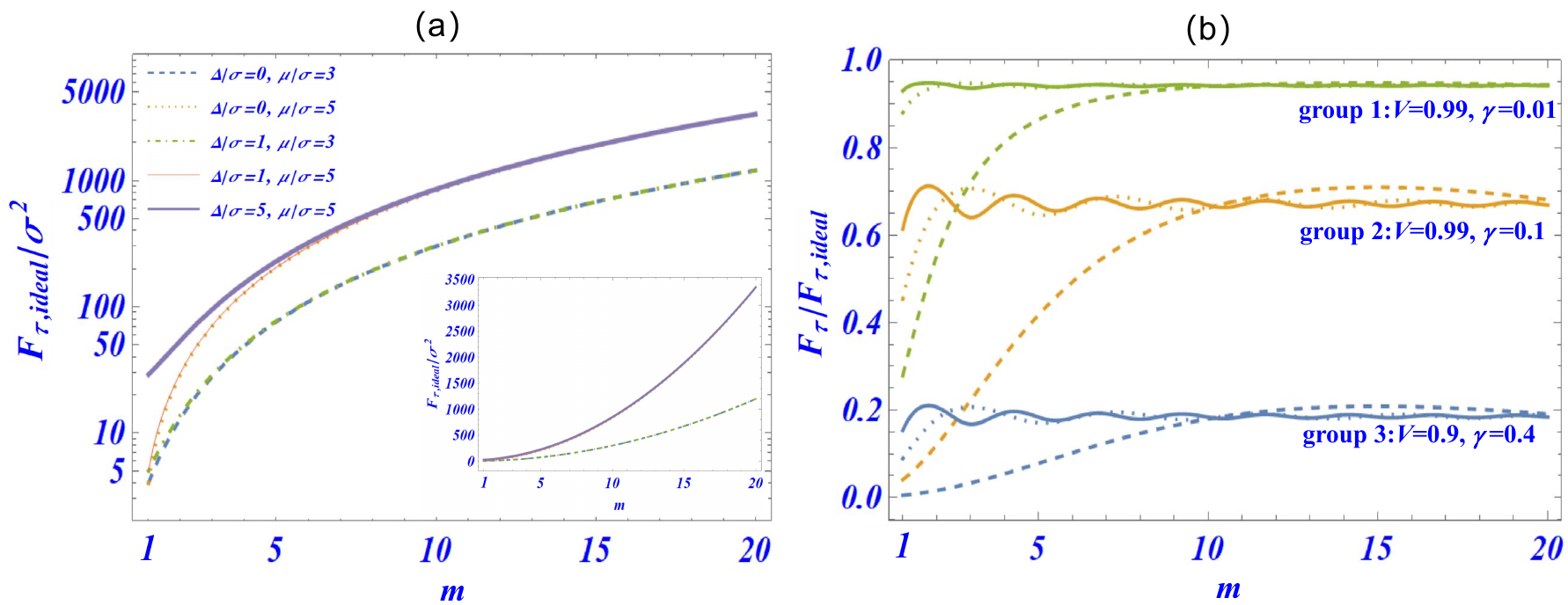}
}}}
\end{picture}
\caption{\label{Fig1}
(a) Fisher information, in units of $\sigma^2$, as a function of mode number $m$ at $\tau \to 0$ in the case of zero loss and perfect visibility ($\gamma=0,V=1$). The horizontal axis of this plot is logarithmic. Fisher information increases significantly as the mode number increases, indicating that the use of BFC as a probe state is an optimal strategy for improving the timing precision. Different curves correspond to different detuning $\Delta$ and mode spacing $\mu$ in units of $\sigma$, which only determine the baseline (initial value) of the precision. The inset shows that Fisher information scales quadratically with $m$, and the parabolic curvatures are determined solely by the mode spacing $\mu$. (b) The ratio $F_{\tau}/F_{\tau,ideal}$, as a function of $m$ for group 1 ($V=0.99,\gamma=0.01$), group 2 ($\gamma=0.1,V=0.99$), and group 3 ($\gamma=0.4,V=0.9$) at delays $\sigma\tau$=0.1 (dashed), 0.5 (dotted), and 0.85 (solid), respectively. The ratios rapidly increase to their maximums (0.94, 0.67, 0.18) with the increase of $m$, especially for the time delays close to zero, indicating that increasing the mode number is also optimal for improving the timing precision in practical scenarios. }
\end{figure}
In the case of zero loss and perfect visibility we recover the QCRB, thus confirming that the measurement strategy is indeed optimal. Fig.\ref{Fig1}(a) shows the FI in units of $\sigma^2$ as a function of mode number $m$ at $\tau \to 0$ under conditions of zero loss and perfect visibility ($\gamma=0,V=1$) for different detuning $\Delta$ and mode spacing $\mu$. It can be seen that FI rapidly increases with the increase of mode number, indicating that increasing mode number is an effective strategy for improving the timing precision. Different detuning $\Delta$ and mode spacing $\mu$ only affect the baseline (initial value) of the precision. The inset in Fig.\ref{Fig1}(a) illustrates a quadratic dependence of FI with respect to mode number $m$, and the parabolic curvatures are determined solely by the mode spacing $\mu$, meaning that the FI increases faster for a larger mode spacing.

To compare FI under ideal and non-ideal conditions, we define a ratio, $F_{\tau}/F_{\tau,ideal}$, to quantify the effects of photon loss $\gamma$ and visibility $V$. The results are plotted in Fig.\ref{Fig1}(b). We find that the ratios equal one in the case of zero loss and perfect visibility, indicating that the quantum precision limit can be achieved under ideal conditions. However, the ratios are always below one when there is loss and imperfect visibility, indicating that the limit cannot be achieved in practice. Despite this, the ratios increase significantly up to their maximums (0.94, 0.67, 0.18) as the mode number $m$ increases, regardless of the photon loss $\gamma$ and visibility $V$. For example, these ratios are obtained in groups with different conditions:  group 1 ($V=0.99,\gamma=0.01$), group 2 ($\gamma=0.1,V=0.99$), and group 3 ($\gamma=0.4,V=0.9$). The effect becomes more pronounced as the time delays approach zero (dashed lines). This indicates that increasing the mode number is also optimal for improving the timing precision in practical scenarios. The maximal precision of HOM interferometry under non-perfect conditions can be calculated by setting $\tau \to \infty$  or $m \to \infty$ in Eq.(\ref{FI-1}). For the time delays that exceed the coherence time of the biphotons, such as $\sigma\tau$=0.85 (solid) in Fig.\ref{Fig1}(b), the FI approaches its maximums regardless of the mode number $m$, but this situation is beyond the scope of our interest. 

Another widely used analytical technique for estimator of $\tau$ is maximum likelihood estimation (MLE) \cite{SA2018,npj2019,CHENPRAPP2023}, where the likelihood function $\mathcal{L}(\tau)$ is defined from measurement outcomes. The logarithm of the likelihood function can be maximized using optimization algorithms such as gradient descent to predict the parameter $\tau$. The likelihood function is a multinomial distribution $\mathcal{L}=R_0^{N_0}R_1^{N_1}R_2^{N_2}$ which can be extremized as \cite{SA2018,npj2019,CHENPRAPP2023}
\begin{eqnarray}
\label{MLE1}
0=:(\partial_{\tau}\log\mathcal{L})_{\tilde\tau_{MLE}}=N_2\frac{\partial_{\tau}R_2(\tau)}{R_2(\tau)}+N_1\frac{\partial_{\tau}R_1(\tau)}{R_1(\tau)}+N_0\frac{\partial_{\tau}R_0(\tau)}{R_0(\tau)}.
\end{eqnarray}
where $N_0$, $N_1$ and $N_2$ denote the numbers of events that no, only one and two detector(s) click(s), respectively. Because $R_0(\tau)$ is a constant and independent of $\tau$, its derivative equals to zero. Also, we find $\partial_{\tau}R_2(\tau)=-\partial_{\tau}R_1(\tau)$, we thus have
\begin{eqnarray}
\label{MLE1-1}
N_2R_1(\tau)=N_1R_2(\tau).
\end{eqnarray}
Solving this equation allows us to predict an optimal estimator, denoted as $\tilde\tau_{MLE}$. Note that the solution of Eq. (\ref{MLE1-1}) can only be found numerically due to the complexity of coincidence probability in Eq. (\ref{R123}). Analytical solutions for Eq. (\ref{MLE1-1}) are feasible only in certain simplified scenarios, such as when $m=1$, as demonstrated in \cite{SA2018,npj2019}.

\section{\label{sec:4} Fisher information of spectrally resolved HOM interferometry with biphoton frequency comb}
Now let us consider whether the ultimate limit in Eq.(\ref{QCR}) can be saturated using spectrally resolved measurement \cite{SR-PRA2015,SR-OE2015,jin2016,PRAPPLIED2020,SR-PR2020,OLT2023,JIN-2024} in HOM interferometry with the probe state given in Eq.(\ref{psi-tau}). We now need to calculate the FI for spectrally resolved HOM interferometry. To this end, the corresponding probability distributions can be rewritten as
\begin{eqnarray}
\label{R123-SR}
&&R_2'(\tau)=\frac{(1-\gamma)^2}{2\sqrt{2 \pi \sigma^2}}\frac{1}{m}\Big(1+V\cos[(\Delta+2\Omega)\tau]\Big)\sum_{k=1}^m e^{-\frac{((2k-m-1)\mu/2-\Omega)^2}{2\sigma^2}},
\nonumber\\
&&R_1'(\tau)=\frac{(1-\gamma)^2}{2\sqrt{2 \pi \sigma^2}}\frac{1}{m}\Big(\frac{1+3\gamma}{1-\gamma}-V\cos[(\Delta+2\Omega)\tau]\Big)\sum_{k=1}^m e^{-\frac{((2k-m-1)\mu/2-\Omega)^2}{2\sigma^2}},\nonumber\\
&&R_0'(\tau)=\frac{\gamma^2}{\sqrt{2 \pi \sigma^2}}\frac{1}{m}\sum_{k=1}^m e^{-\frac{((2k-m-1)\mu/2-\Omega)^2}{2\sigma^2}}.
\end{eqnarray}
\begin{figure}[th]
\begin{picture}(380,430)
\put(0,0){\makebox(375,430){
\scalebox{0.8}[0.8]{
\includegraphics{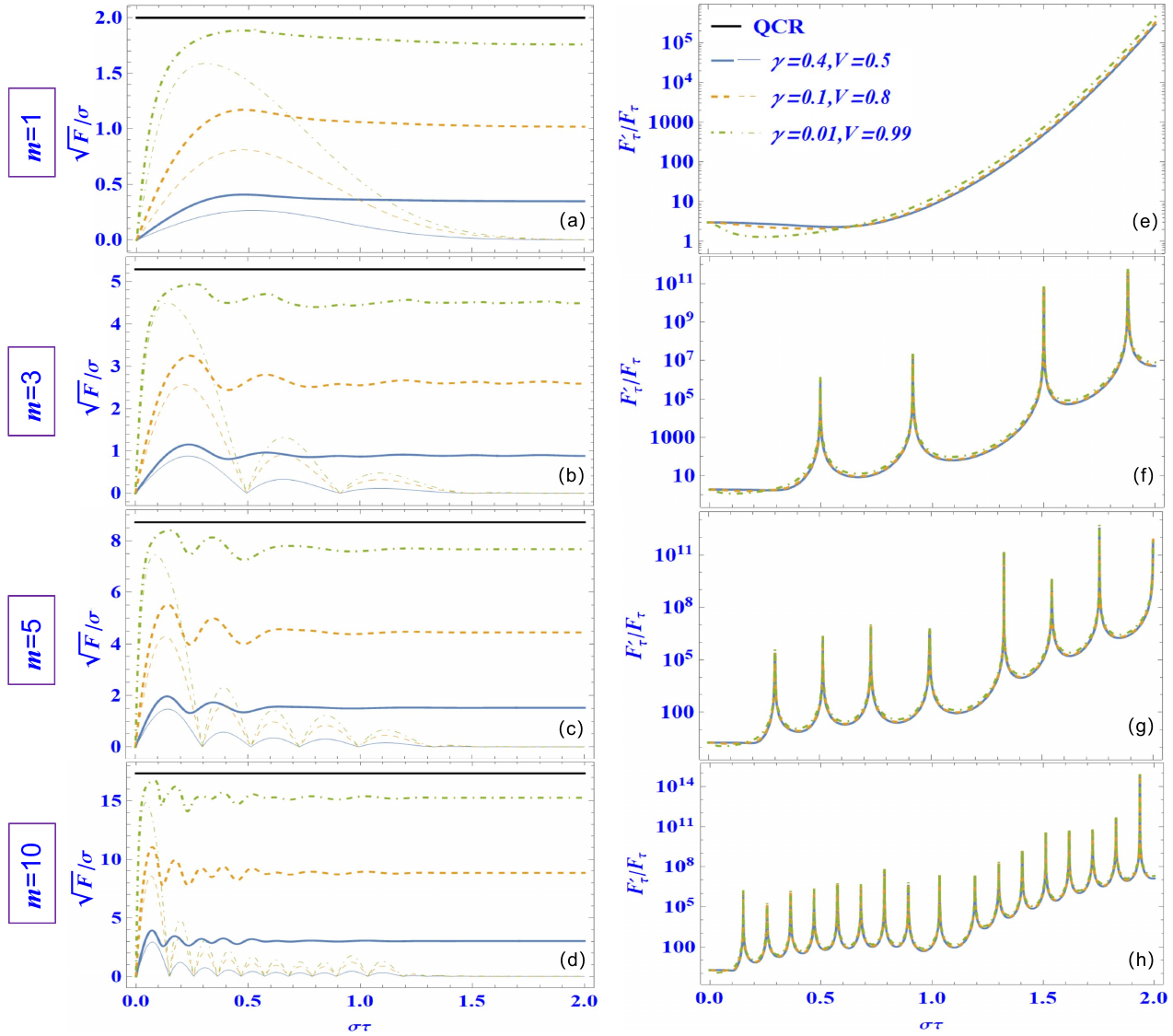}
}}}
\end{picture}
\caption{\label{Fig2}
Left column: The square root of Fisher information in units of $\sigma$ as a function of $\sigma\tau$ for $m$=1,3,5,10 from top to bottom, respectively. All values of Fisher information are below the QCRB (black bold lines). Spectrally resolved results (thick lines) are always superior to spectrally non-resolved ones (thin lines) in all cases. The blue, orange, and green curves represent $\gamma=0.4,V=0.5$, $\gamma=0.1,V=0.8$, and $\gamma=0.01,V=0.99$, respectively. We set the detuning $\Delta/\sigma=0$ and mode spacing $\mu/\sigma=3$ for all plots. Plots in the right column are corresponding enhancement factor $F'_{\tau}/F_{\tau}$.}
\end{figure}
then, the FI can be rewritten as \cite{CHENPRAPP2023,PRAPP2023}
\begin{equation}
\label{FI-SR}
F'_{\tau}=\int \Big(\frac{(\partial_{\tau}R'_2(\tau))^2}{R'_2(\tau)}+\frac{(\partial_{\tau}R'_1(\tau))^2}{R'_1(\tau)}+\frac{(\partial_{\tau}R'_0(\tau))^2}{R'_0(\tau)}\Big)d\Omega .
\end{equation}
Substituting Eq.(\ref{R123-SR}) into Eq.(\ref{FI-SR}), we finally arrive at
\begin{eqnarray}
\label{FI-SR-1}
F'_{\tau}=\int\frac{V^2(1-\gamma)^2(1+\gamma)(\Delta+2\Omega)^2\sin[(\Delta+2\Omega)\tau]^2\sum_{k=1}^m e^{-\frac{((2k-m-1)\mu/2-\Omega)^2}{2\sigma^2}}}{m\sqrt{2 \pi \sigma^2}(V\cos[(\Delta+2\Omega)\tau]+1)(1+3\gamma-V(1-\gamma))(\cos[(\Delta+2\Omega)\tau])}d\Omega.
\end{eqnarray}
In the case of zero loss and perfect visibility, i.e., $\gamma=0,V=1$, we have
\begin{eqnarray}
\label{FI-SR-11}
F'_{\tau,ideal}&&=\int\frac{(\Delta+2\Omega)^2\sum_{k=1}^m e^{-\frac{((2k-m-1)\mu/2-\Omega)^2}{2\sigma^2}}}{m\sqrt{2 \pi \sigma^2}}d\Omega \nonumber\\
&&=\sum_{k=1}^m \int\frac{(\Delta+2\Omega)^2 e^{-\frac{((2k-m-1)\mu/2-\Omega)^2}{2\sigma^2}}}{m\sqrt{2 \pi \sigma^2}}d\Omega=(m^2-1)\mu^2/3+\Delta^2+4\sigma^2.
\end{eqnarray}

From Eq.(\ref{FI-SR-11}), it can be seen that we recover the QCRB in the case of zero loss and perfect visibility, thus confirming that the measurement strategy is indeed optimal. It is worth noting that under the ideal condition of zero loss and perfect visibility, the FI with spectrally resolved measurement does not depend on the time delay $\tau$. This means that one can acheive the QCRB at any time delays, even when the time delay exceeds the coherence time of the biphotons, referred to as the ambiguity-free dynamic range in \cite{CHENPRAPP2023}. However, this is not the case with spectrally non-resolved measurement, where the FI always depends on the time delay even under the ideal condition of zero loss and perfect visibility (see Eq.(\ref{FI-ideal})). In this case, the QCRB can realized only at zero delay. This represents an advantage of spectrally resolved scheme compared to spectrally non-resolved one, as demonstrated in \cite{CHENPRAPP2023} for $m=1$.

Analogous to the end of Part III, the likelihood function for spectrally resolved scheme is $\mathcal{L}=R'{_0^{N_0}}R'{_1^{N_1}}R'{_2^{N_2}}$, which can be extremized as 
\cite{SA2018,npj2019,CHENPRAPP2023}
\begin{eqnarray}
\label{MLE2}
0=:(\partial_{\tau}\log\mathcal{L})_{\tilde\tau_{MLE}}=\sum_{\Omega}N_2\frac{\partial_{\tau}R'_2(\tau)}{R'_2(\tau)}+N_1\frac{\partial_{\tau}R'_1(\tau)}{R'_1(\tau)}+N_0\frac{\partial_{\tau}R'_0(\tau)}{R'_0(\tau)}.
\end{eqnarray} 
Here $R'_0(\tau)$ is independent of the parameter of interest $\tau$, and thus it can be neglected. Based on the relation $\partial_{\tau}R'_2(\tau)=-\partial_{\tau}R'_1(\tau)$, Eq. (\ref{MLE2}) can be simplified as
\begin{eqnarray}
\label{MLE2-1}
\sum_{\Omega}N_2R'_1(\tau)=\sum_{\Omega}N_1R'_2(\tau).
\end{eqnarray}
Solving this equation enables us to predict an optimal estimator as $\tilde\tau_{MLE}$. Analogously,  the solution of Eq. (\ref{MLE2-1}) can only be found numerically because of the complexity of the coincidence probability in Eq. (\ref{R123-SR}). The analytical solution of Eq. (\ref{MLE2-1}) can be achieved only in some simplified cases, e.g., for $m=1$, as shown in \cite{CHENPRAPP2023}.

\section{\label{sec:5}Comparison of Fisher information between spectrally non-resolved and resolved HOM interferometry under imperfect conditions}

To compare the advantage of the spectrally resolved scheme with the spectrally non-resolved one, we define an enhancement factor, $F'_{\tau}/F_{\tau}$, for non-ideal conditions ($\gamma\neq0,V\neq1$). The results are plotted in Fig. \ref{Fig2}, where left column represents the square root of FI in units of $\sigma$ as a function of $\sigma\tau$ for $m$=1, 3, 5, 10 from top (a) to bottom (d), respectively. It can be seen that all values of the FI are below the QCRB (black bold lines) under non-ideal conditions, while the spectrally resolved results (thick lines) are close to the QCRB and superior to the spectrally non-resolved ones (thin lines) in all cases. Additionally, the FI increases as the mode number $m$ increases while the maximum of the FI is not at zero delay but is shifted to other delay points. The curves of FI for the spectrally non-resolved scheme exhibit one, three, five, and more than ten peaks for $m=1,3,5,10,$ but they ultimately tend to zero as the time delay increases. This is due to the fact that, with the increase of the time delay, it will exceed the scope of the coherence time of the biphotons, where the timing based on spectrally non-resolved interference is unavailable. On the other hand, the oscillations that appeare for multiple frequency modes are related to the modulations in the temporal HOM interferogram introduced by mode number $m$ and spacing $\mu$, as described in the end of Part II. In contrast, the curves of FI for the spectrally resolved scheme exhibit fewer oscillations and rapidly tend to a constant as the time delay increases. This indicate that spectrally resolved scheme remains valid even if the time delay exceeds the scope of biphoton coherence time, which is a significant advantage compared to the spectrally non-resolved scheme.

To quantify this advantage, we plot the enhancement factor $F'_{\tau}/F_{\tau}$ in the right column in Fig. \ref{Fig2}. For a single mode $m=1$, the enhancement factor exhibits an exponential increase with the increasing of the time delay in Fig. \ref{Fig2}(e). This result is somewhat similar to that  reported in \cite{PRAPP2023} with two independent photons. For $m=3, 5, 10$, as shown in Fig. \ref{Fig2}(f)-(h), the enhancement factors continue to grow with the increase of $m$ for a fixed delay, indicating that the enhancement is more pronounced for lagre $m$. Additionally, it exhibits more oscillations as $m$ increases due to the additional oscillations introduced by mode number $m$ and spacing $\mu$.

\section{\label{sec:conclude}discussion and CONCLUSIONS}

Based on the anysis above, we find that the QCRB can be acheived by HOM interferometry in both spectrally resolved and non-resolved manners under ideal conditions of zero losses and perfect visibility. The spectrally resolved scheme is a better strategy because it offers higher Fisher information, shorter measurement times and ambiguity-free dynamic range. In practice, however, these conditions cannot be satisfied, so the precision never reaches the quantum limit. Nevertheless, one can approach the practical precision limits faster by adopting the strategy of increasing the mode number of BEC. Furthermore, one can also approach the precision limits under realistic circumstances by optimizing some experimental parameter, such as visibility and the shape of the JSA, as shown in \cite{PRL2024} recently. Additionally, obtaining the optimal estimator of $\tilde\tau_{MLE}$ using MLE method for the BEC probe state is challenging due to the complexity of coincidence probability involved.

Note that the time delay is introduced only in one of the arms of the HOM interferometer, meaning that the dynamical evolution is implemented on only one photon of the photon pair. However, it is possible to implement the dynamical evolution for both photons by adding a delay $-\tau$ in other arm of the interferometer. This adjustment allows us to attain a phase factor multiplied by 2, resulting in an additional precision improvement of 2.

Experimentaly, the BFC can be prepared with various ways, including a resonant cavity \cite{PRL2003,NP2015,Optica2017-,OE-2018,npj2021,CP2023}, engineered nonlinear crystal \cite{PRA2012,OE2019-2,APLP2022}, filtering (pulse shaping) \cite{NJP2014,OE2019,PRL2020}, or spectrally resolved technology \cite{jin2016,NPJ2021,OLT2023}. These preparations can be performed either in bulk nonlinear crystals,  chips \cite{OE-2018,Nature2017,Optica2017-}, or integrated photonic waveguide devices \cite{Fabre2020}. Furthermore, it is relatively convenient to manipulate the mode number of BFC, which can be extremely large \cite{npj2021,SRP2022}. Of course, for a fixed mode spacing, increaseing the mode number also broadens the spectrm of frequency comb. However, the temporal envelope of a multi-mode HOM interferogram remains unaffected, as it is determined solely  by the width of a single frequency mode.

HOM interferometry with multiple frequency mode can also be used in cases involving extremely sensible samples, achieving attosecond-scale (nanometer path length) resolution, such as in biological sensing \cite{PR2016}. Our analysis in this paper can also be extended to multi-mode N00N state interferometry \cite{JINOE2023} for exploring the precision limits of phase estimation in a similar manner.

In conclusion, we examine the ultimate quantum timing limits using the BFC as a probe state and obtain a QCRB that scales quadratically with the number of frequency modes. This provides the ultimate sensitivity allowed by quantum physics in the estimation of the time delay between two photons by measuring their interference at a beam splitter. This bound can be saturated by spectrally non-resolved HOM interferometry at zero delay, and by spectrally resolved HOM interferometry at arbitrary delays under ideal conditions. The FI is below the bound for imperfect experimental conditions, but it rapidly increases up to its maximum as the  mode number increases for a fixed time delay close to zero, indicating that increasing the mode number is an optimal strategy for improving the timing precision in practice. Furthermore, compared with sepctral non-resolved measurements, spectrally resolved measurements are a better strategy due to their higher Fisher information, shorter measurement times and ambiguity-free dynamic range.  This work offers a theoretical timing limit that can be achieved using the BFC as a probe state and an optimal strategy for improving the timing precision of HOM-based quantum metrology in practical scenarios.

\begin{acknowledgments}
This work has been supported by National Natural Science Foundation of China (12074309, 12074299, 12033007, 61875205, 12103058, 61801458), the Youth Innovation Team of Shaanxi Universities, and the Natural Science Foundation of Hubei Province (2022CFA039).
\end{acknowledgments}

\appendix
\section{The derivation of QCRB using BFC state as a probe}
The QCRB on the estimation of time delays can be calculated using Eqs. (\ref{delta}) and (\ref{Q}) with a BFC probe state in Eq. (\ref{psi-tau}). The first term in Eq. (\ref{Q}) can be calculated as follows,
\begin{eqnarray}
\label{Q1}
&& \Big\langle\frac{\partial\Psi(\tau)}{\partial\tau}\Big|\frac{\partial\Psi(\tau)}{\partial\tau}\Big\rangle\nonumber\\
&&=\frac{1}{2}\Big(\langle 0| \sum_m\int d\Omega (-i(\omega_{s0}+\Omega))f^*(\Omega+m\mu)e^{-i(\omega_{s0}+\Omega)\tau}\hat{a}_{1}(\omega_{s0}+\Omega)\hat{a}_{2}(\omega_{i0}-\Omega)
\nonumber\\
&& +\sum_m\int d\Omega (-i(\omega_{i0}-\Omega))f^*(\Omega+m\mu)e^{-i(\omega_{i0}-\Omega)\tau}\hat{a}_{1}(\omega_{i0}-\Omega)\hat{a}_{2}(\omega_{s0}+\Omega)\Big) \nonumber\\
&&\times \Big( \sum_m\int d\Omega(i(\omega_{s0}+\Omega))f(\Omega+m\mu)e^{i(\omega_{s0}+\Omega)\tau}\hat{a}_{1}^\dag(\omega_{s0}+\Omega)\hat{a}_{2}^\dag(\omega_{i0}-\Omega)\nonumber\\
&&+\sum_m\int d\Omega(i(\omega_{i0}-\Omega))f(\Omega+m\mu)e^{i(\omega_{i0}-\Omega)\tau}\hat{a}_{1}^\dag(\omega_{i0}-\Omega)\hat{a}_{2}^\dag(\omega_{s0}+\Omega)|0\rangle
 \Big)\nonumber\\
&&=\frac{1}{2} \sum_m\int d\Omega \Big((\omega_{s0}+\Omega)^2+(\omega_{i0}-\Omega)^2\Big)|f(\Omega+m\mu)|^2 \Big( \langle 0|\hat{a}_{1}(\omega_{s0}+\Omega)\hat{a}_{1}^\dag(\omega_{s0}+\Omega)\hat{a}_{2}(\omega_{i0}-\Omega)\hat{a}_{2}^\dag(\omega_{i0}-\Omega)|0\rangle
 \Big)\nonumber\\
&&=\frac{1}{2} \sum_m\int d\Omega \Big(\omega_{s0}^2+\omega_{i0}^2+2(\omega_{s0}-\omega_{i0})\Omega+2\Omega^2\Big)|f(\Omega+m\mu)|^2 
\end{eqnarray}
In above derivation, we use the relations $\hat{a}(\omega_s)\hat{a}^\dag(\omega_i)=\delta(\omega_s-\omega_i)$ and $\langle 0|0\rangle=1$. From the last line in Eq.(\ref{Q1}) we can see that it denotes taking the expectation value of the four terms in big bracket with respect to the function of $|f|^2$. 
Taking the JSA as a Gaussian function, i.e., $f(\Omega+m\mu)= \exp(-((2k-m-1)\mu/2+\Omega)^2/4\sigma^2)$, and using the normalized condition $\sum_m\int_{-\infty}^{\infty}|f(\Omega+m\mu)|^2=1$, we have 
\begin{eqnarray}
\label{Q1-2}
 \Big\langle\frac{\partial\Psi(\tau)}{\partial\tau}\Big|\frac{\partial\Psi(\tau)}{\partial\tau}\Big\rangle
&&=\frac{1}{2}(\omega_{s0}^2+\omega_{i0}^2)+(\omega_{s0}-\omega_{i0})\langle\Omega\rangle+\langle\Omega^2\rangle\nonumber\\
&&=\frac{1}{2}(\omega_{s0}^2+\omega_{i0}^2)+\frac{1}{m C}\sum_m \Big((2k-m-1)^2\mu^2 /2+2\sigma^2\Big)\nonumber\\
&&=\frac{1}{2}(\omega_{s0}^2+\omega_{i0}^2)+(m^2-1)\mu^2/12+\sigma^2
\end{eqnarray}
where $C$ is a constant associated with the normalized factor. In above derivation, the expectation value of $\Omega$ equals to zero, i.e., $\langle\Omega\rangle=0$. The second term in Eq. (\ref{Q}) can be calculated in a similar manner,
\begin{eqnarray}
\label{Q2}
&& \Big|\Big\langle \Psi(\tau) \Big|\frac{\partial\Psi(\tau)}{\partial\tau}\Big\rangle\Big|^2\nonumber\\ &&=\Big|\frac{1}{2} \langle 0|\sum_m\int d\Omega |f(\Omega+m\mu)|^2(i(\omega_{s0}+\Omega+\omega_{i0}-\Omega)\hat{a}_{1}(\omega_{s0}+\Omega)\hat{a}_{1}^\dag(\omega_{s0}+\Omega)\hat{a}_{2}(\omega_{i0}-\Omega)\hat{a}_{2}^\dag(\omega_{i0}-\Omega)|0\rangle \Big|^2
\nonumber\\&&=\Big|\frac{i}{2} \sum_m\int d\Omega (\omega_{s0}+\omega_{i0}) |f(\Omega+m\mu)|^2 \Big|^2
\nonumber\\&&=\frac{1}{4}\Big( \sum_m\int d\Omega (\omega_{s0}+\omega_{i0})|f(\Omega+m\mu)|^2 \Big)^2
\end{eqnarray}
The last line in big bracket in Eq.(\ref{Q2}) denotes taking the expectation value of $\omega_{s0}+\omega_{i0}$  with respect to the function of $|f|^2$. Eq.(\ref{Q2}) can thus be simplified as,
\begin{eqnarray}
\label{Q2-}
\Big|\Big\langle \Psi(\tau) \Big|\frac{\partial\Psi(\tau)}{\partial\tau}\Big\rangle\Big|^2=\frac{1}{4}(\omega_{s0}+\omega_{i0})^2
\end{eqnarray}
Combine Eqs.(\ref{Q1-2}) and (\ref{Q2-}), we have 
\begin{eqnarray}
\label{Q'}
Q=&&4\Big(\Big\langle\frac{\partial\Psi(\tau)}{\partial\tau}\Big|\frac{\partial\Psi(\tau)}{\partial\tau}\Big\rangle-\Big|\Big\langle \Psi(\tau) \Big|\frac{\partial\Psi(\tau)}{\partial\tau}\Big\rangle\Big|^2\Big)\nonumber\\&&=4\Big(\frac{1}{2}(\omega_{s0}^2+\omega_{i0}^2)+(m^2-1)\mu^2/12+\sigma^2-\frac{1}{4}(\omega_{s0}+\omega_{i0})^2 \Big)\nonumber\\&&=(m^2-1)\mu^2/3+\Delta^2+4\sigma^2.
\end{eqnarray}
where $\Delta=\omega_{s0}-\omega_{i0}$. We thus obtain the QCRB (Eq. (\ref{QCR})) in the main text. 

\section{The derivation of coincidence probability for HOM interferometry}
In this Section, we give the derivation of coincidence probability for HOM interferometry from the frequency domain. The two-photon state, for example, generated from a spontaneous parametric down-conversion (SPDC) process, can be described as
\begin{equation}
|\Psi\rangle =\int\int d\omega_s d\omega_i f(\omega_s, \omega_i)\hat{a}_s^\dag(\omega_s)\hat{a}_i^\dag(\omega_i)|0\rangle
\end{equation}
where $\omega$ is the angular frequency, and $\hat{a}_{s,i}^\dag$ is the creation operator and the subscripts $s$ and $i$ denote the signal and idler photons from SPDC, respectively. $|0\rangle$ stands for a vacuum state. $f(\omega_s, \omega_i)$ is the JSA. We use this probe state to interact with dynamic system, i.e., introduce a relative time delay $\tau$ in the path of idler photon, it implies that a relative phase shift of $e^{-i\omega_2\tau}$ is added. The transformation rule of the 50/50 beamsplitter (BS) is
\begin{eqnarray}
\hat{a}_1(\omega_1)= [\hat{a}_{s}(\omega_1)+\hat{a}_i(\omega_1)e^{-i\omega_1\tau}]/\sqrt{2},\nonumber\\
\hat{a}_2(\omega_2)= [\hat{a}_{s}(\omega_2)-\hat{a}_i(\omega_2)e^{-i\omega_2\tau}]/\sqrt{2}.
\end{eqnarray}
The detection field operators of two detectors used for the coincidence count rates are
\begin{eqnarray}
\hat{E}_1^{(+)}(t_1)= \frac{1}{\sqrt{2\pi}}\int_{0}^{\infty}d\omega_1\hat{a}_1(\omega_1)e^{-i\omega_1t_1},\nonumber\\
\hat{E}_2^{(+)}(t_2)= \frac{1}{\sqrt{2\pi}}\int_{0}^{\infty}d\omega_2\hat{a}_2(\omega_2)e^{-i\omega_2t_2}.
\end{eqnarray}
where the subscripts 1 and 2 denote the photons detected by D1 and D2, respectively. Then, the coincidence count rates between two detectors as a function of time delay $\tau$ can be expressed as
\begin{equation}
R(\tau)= \int \int d t_1 d t_2 \langle \Psi |\hat{E}_1^{(-)}\hat{E}_2^{(-)} \hat{E}_2^{(+)}\hat{E}_1^{(+)}|\Psi \rangle=\int \int d t_1 d t_2 |\langle 0| \hat{E}_2^{(+)}\hat{E}_1^{(+)}|\Psi \rangle|^2
\end{equation}
Consider $ \hat{E}_2^{(+)}\hat{E}_1^{(+)}|\Psi \rangle$, only 2 out of 4 terms exist. The first term is
\begin{eqnarray}
&&-\frac{1}{2\pi}\int_{0}^{\infty}\int_{0}^{\infty}d\omega_1d\omega_2\hat{a}_s(\omega_1)\hat{a}_i(\omega_2)e^{-i\omega_2 \tau} e^{-i\omega_1 t_1}e^{-i\omega_2 t_2} \int_{0}^{\infty}\int_{0}^{\infty} d\omega_s d\omega_i f(\omega_s, \omega_i)\times \hat{a}_s^\dag(\omega_s)\hat{a}_i^\dag(\omega_i)|0\rangle \nonumber\\
&&= -\frac{1}{2\pi}\int_{0}^{\infty}\int_{0}^{\infty} d\omega_1 d\omega_2 f(\omega_1, \omega_2)e^{-i\omega_1 t_1}e^{-i\omega_2 t_2}e^{-i\omega_2\tau}|0\rangle
\end{eqnarray}
In this calculation, the relationship of $\hat{a}_s(\omega_1)\hat{a}_s^\dag(\omega_s)=\delta(\omega_1-\omega_s),\hat{a}_i(\omega_2)\hat{a}_i^\dag(\omega_i)=\delta(\omega_2-\omega_i)$ are used.

The second term is
\begin{eqnarray}
&&\frac{1}{2\pi}\int_{0}^{\infty}\int_{0}^{\infty}d\omega_1d\omega_2\hat{a}_i(\omega_1)\hat{a}_s(\omega_2)e^{-i\omega_1 \tau} e^{-i\omega_2 t_2}e^{-i\omega_1 t_1} \int_{0}^{\infty}\int_{0}^{\infty} d\omega_s d\omega_i f(\omega_s, \omega_i)
 \hat{a}_s^\dag(\omega_s)\hat{a}_i^\dag(\omega_i)|0\rangle\nonumber\\
&&= \frac{1}{2\pi}\int_{0}^{\infty}\int_{0}^{\infty} d\omega_1 d\omega_2 f(\omega_2, \omega_1)e^{-i\omega_2 t_2}e^{-i\omega_1 t_1}e^{-i\omega_1\tau}|0\rangle
\end{eqnarray}
Combine these two terms:
\begin{eqnarray}
\hat{E}_2^{(+)}\hat{E}_1^{(+)}|\Psi \rangle=&&\frac{1}{2\pi}\int_{0}^{\infty}\int_{0}^{\infty}d\omega_1d\omega_2e^{-i\omega_1 t_1}e^{-i\omega_2 t_2}\times[f(\omega_2, \omega_1)e^{-i\omega_1 \tau}-f(\omega_1, \omega_2)e^{-i\omega_2 \tau}]|0\rangle
\end{eqnarray}
Then,
\begin{eqnarray}
\langle \Psi |\hat{E}_1^{(-)}\hat{E}_2^{(-)} \hat{E}_2^{(+)}\hat{E}_1^{(+)}|\Psi \rangle=\left(\frac{1}{2\pi}\right)^2\int_{0}^{\infty}\int_{0}^{\infty}d\omega_1d\omega_2d\omega_1^{'}d\omega_2^{'}e^{-i(\omega_1-\omega_1^{'}) t_1}e^{-i(\omega_2-\omega_2^{'}) t_2}\nonumber\\\times[f^{*}(\omega_2^{'}, \omega_1^{'})e^{-i\omega_1^{'} \tau}-f^{*}(\omega_1^{'}, \omega_2^{'})e^{-i\omega_2^{'} \tau}][f(\omega_2, \omega_1)e^{-i\omega_1\tau}-f(\omega_1, \omega_2)e^{-i\omega_2 \tau}]
\end{eqnarray}
Finally,
\begin{eqnarray}
R(\tau)= \int \int d t_1 d t_2 \langle \Psi |\hat{E}_1^{(-)}\hat{E}_2^{(-)} \hat{E}_2^{(+)}\hat{E}_1^{(+)}|\Psi \rangle=\frac{1}{2}\int_{0}^{\infty}\int_{0}^{\infty}d\omega_1d\omega_2d\omega_1^{'}d\omega_2^{'}\delta(\omega_1-\omega_1^{'})\delta(\omega_2-\omega_2^{'})\nonumber\\\times
[f^{*}(\omega_2^{'}, \omega_1^{'})e^{-i\omega_1^{'} \tau}-f^{*}(\omega_1^{'}, \omega_2^{'})e^{-i\omega_2^{'} \tau}][f(\omega_2, \omega_1)e^{-i\omega_1 \tau}-f(\omega_1, \omega_2)e^{-i\omega_2 \tau}]
\end{eqnarray}
In above calculation, the relationship of $\delta(\omega-\omega^{'})=\frac{1}{2\pi}\int_{-\infty}^{\infty}e^{i(\omega-\omega^{'})t}dt$ is used. $f^*$ is the complex conjugate of $f$. 
\begin{eqnarray}
\label{R12}
R(\tau)= \frac{1}{4}\int_{0}^{\infty}\int_{0}^{\infty}d\omega_1d\omega_2|f(\omega_1, \omega_2)e^{-i\omega_2 \tau}-f(\omega_2, \omega_1)e^{-i\omega_1 \tau}|^2
\end{eqnarray}
In order to introduce less variables, Eq.(\ref{R12}) can be rewritten as
\begin{eqnarray}
\label{R}
R(\tau)=&& \frac{1}{4}\int_{0}^{\infty}\int_{0}^{\infty}d\omega_sd\omega_i|f(\omega_s, \omega_i)e^{-i\omega_i \tau}-f(\omega_i, \omega_s)e^{-i\omega_s \tau}|^2\nonumber\\
=&&\frac{1}{4}\int_{0}^{\infty}\int_{0}^{\infty}d\omega_sd\omega_i|f(\omega_s, \omega_i)-f(\omega_i, \omega_s)e^{-i(\omega_s-\omega_i) \tau}|^2\nonumber\\
=&&\frac{1}{4}\int_{0}^{\infty}\int_{0}^{\infty}d\omega_sd\omega_i\Big(|f(\omega_s, \omega_i)|^2+|f(\omega_i, \omega_s)|^2-2f(\omega_s, \omega_i)f^*(\omega_i, \omega_s)\cos[(\omega_s-\omega_i) \tau]\Big)
\end{eqnarray}
For simplicity, we only consider the JSA to be real and normalized, i.e., $f(\omega_s, \omega_i)=f^*(\omega_s, \omega_i)$ and $\int \int d\omega_s d\omega_i |f(\omega_s,\omega_i)|^2$=1, then 
\begin{eqnarray}
\label{Rsi}
R(\tau)=\frac{1}{2}-\frac{1}{2}\int_{0}^{\infty}\int_{0}^{\infty}d\omega_sd\omega_i\Big(f(\omega_s, \omega_i)f^*(\omega_i, \omega_s)\cos[(\omega_s-\omega_i) \tau]\Big)
\end{eqnarray}
For the CW-pumped SPDC source, we have the relations $\omega_{s}=\omega_{s0}+\Omega, \omega_{i}=\omega_{i0}-\Omega$, and the JSA in variables $\omega_s,\omega_i$ can be changed into a single one $\Omega$, then two-photon state becomes
\begin{equation}
|\Psi\rangle =\frac{1}{\sqrt{2}}\int d\Omega f(\Omega)\Big(\hat{a}_1^\dag(\omega_{s0}+\Omega)\hat{a}_2^\dag(\omega_{i0}-\Omega)+e^{i\phi} \hat{a}_1^\dag(\omega_{i0}-\Omega)\hat{a}_2^\dag(\omega_{s0}+\Omega)\Big)|0\rangle
\end{equation}
The above equation represents a frequency entangled state \cite{Cyril2023,PRL2009}, and $\phi$ is the relative phase between two substates. Then, the coincidence probability becomes,
\begin{eqnarray}
\label{Rsi-}
R(\tau)=\frac{1}{2}-\frac{1}{2}\int_{-\infty}^{\infty}d\Omega\Big(f(\Omega)f^*(-\Omega)\cos[(\Delta+2\Omega)\tau+\phi]\Big)
\end{eqnarray}
Now, we take the JSA as a multi-mode Gaussian function as follows \cite{APLP2022,JINOE2023},
\begin{eqnarray}
\label{f(s,i)}
f(\omega_s, \omega_i)&&=\sum_{k=1}^m e^{-\frac{(\omega_s-\omega_{s0}-(2k-m-1)\mu/2)^2}{8\sigma^2}-\frac{(\omega_i-\omega_{i0}+(2k-m-1)\mu/2)^2}{8\sigma^2}}\nonumber\\
&&=f(\Omega)=\sum_{k=1}^m e^{-\frac{(\Omega-(2k-m-1)\mu/2)^2}{8\sigma^2}-\frac{(-\Omega+(2k-m-1)\mu/2)^2}{8\sigma^2}}\nonumber\\
&&=\sum_{k=1}^m e^{-\frac{((2k-m-1)\mu/2-\Omega)^2}{4\sigma^2}},
\end{eqnarray}
Using $\Delta=\omega_{s0}-\omega_{i0}$ and substituting Eq.(\ref{f(s,i)}) into Eq.(\ref{Rsi-}), we have
\begin{eqnarray}
&&R(\tau)=\frac{1}{2}-\frac{1}{2}\int_{-\infty}^{\infty}d\Omega\sum_{k=1}^m e^{-\frac{((2k-m-1)\mu/2-\Omega)^2}{4\sigma^2}}\times\sum_{k=1}^m e^{-\frac{((2k-m-1)\mu/2+\Omega)^2}{4\sigma^2}}  \cos[(\Delta+2\Omega)\tau+\phi]
\end{eqnarray}
If we assume $\mu\gg\sigma$, the cross terms in above equation can be ignored \cite{JINOE2023}. Then,
\begin{eqnarray}
R(\tau)&&=\frac{1}{2}-\frac{1}{2m}\sum_{k=1}^m\int_{-\infty}^{\infty}d\Omega e^{-\frac{\Omega^2}{2\sigma^2}}  \cos[(\Delta+2\Omega)\tau+\phi+(2k-m-1)\mu\tau]
\nonumber\\
&&=\frac{1}{2}-\frac{1}{2m}\sum_{k=1}^m\int_{-\infty}^{\infty}d\Omega e^{-\frac{\Omega^2}{2\sigma^2}}  \times\Big(\cos[(\Delta+2\Omega)\tau+\phi]\cos[(2k-m-1)\mu\tau]-\sin[(\Delta+2\Omega)\tau+\phi]\sin[(2k-m-1)\mu\tau]\Big)\nonumber\\
&&=\frac{1}{2}-\frac{1}{2m}\sum_{k=1}^m\cos[(2k-m-1)\mu\tau]\int_{-\infty}^{\infty}d\Omega e^{-\frac{\Omega^2}{2\sigma^2}} \cos[(\Delta+2\Omega)\tau+\phi]\nonumber\\
&&=\frac{1}{2}-\frac{1}{2m}\frac{\sin(m\mu\tau)}{\sin(\mu\tau)}\cos(\Delta\tau+\phi) Re \Big[\int_{-\infty}^{\infty}d\Omega  e^{-\frac{\Omega^2}{2\sigma^2}}e^{i2\Omega\tau}\Big]\nonumber\\
&&=\frac{1}{2}-\frac{1}{2m}\frac{\sin(m\mu\tau)}{\sin(\mu\tau)}\cos(\Delta\tau+\phi)e^{-2\sigma^2\tau^2}.
\end{eqnarray}
In above derivation, we use the normalized condition $\int d\Omega |f(\Omega)|^2$=1. $Re$ denotes the real part. After considering the photon loss $\gamma$ and imperfect visibility $V$ \cite{SA2018,npj2019}, we can obtain Eqs. (\ref{R2}) and (\ref{R123}) in the main text.

\end{CJK*}
\end{document}